# The Geometry of Information Cocoon: Analyzing the Cultural Space with Word Embedding Models

Huimin Xu[a], Zhicong Chen[b], Ruiqi Li[c], and Cheng-Jun Wang[a]*

[a] School of Journalism and Communication, Nanjing University, Nanjing, China, 210023

[b] Department of Media and Communication, City University of Hong Kong, Hong Kong, China, 999077

[c] UrbanNet Lab, College of Information Science and Technology, Beijing University of Chemical Technology, Beijing, China, 100029

\* *Correspondence: wangchj04@126.com (C.J. W)，+86 17751010496*

*The word count of the manuscript: 7541*

# The Geometry of Information Cocoon: Analyzing the Cultural Space with Word Embedding Models

Accompanied by the development of digital media, the threat of information cocoon has become a significant issue. However, little is known about the measure of information cocoon as a cultural space and its relationship with social class. This study addresses this problem by constructing the cultural space with word embedding models and random shuffling methods among three large-scale digital media use datasets. In the light of Bourdieu's field theory of cultural production, we investigate the information cocoon effect on different social classes among 979 computer users, 100,000 smartphone users, and 159,373 mobile reading application users. Our analysis reveals that information cocoons widely exist in the daily use of digital media. Moreover, people of lower social class have a higher probability of getting stuck in the information cocoon filled with the entertainment content. In contrast, the people of higher social class have more capability to stride over the constraints of the information cocoon. The results suggest that the disadvantages for vulnerable groups in acquiring knowledge may further widen social inequality.

**Introduction**

Nowadays, the threat of information cocoon is becoming an increasingly severe issue. Although digital media is revolutionizing our daily life by helping people find the personalized information quickly (Qian et al., 2013), it also leads to many concerns about information cocoons (Borgesius et al., 2016). Sunstain (2006) first proposes the term of information cocoon and defines it as "communication universes in which we hear only what we choose and only what comforts us and pleases us" (Sunstein, 2006, p.9). Therefore, information cocoon is essentially a geometric space (Sunstein, 2006; Bourdieu, 1984). There are many other similar concepts, such as the echo chamber (Jamieson & Cappella, 2008), filter bubble (Pariser, 2011), and information enclaves (Weeks et al., 2016). They all emphasize the process of selective exposure to like-minded opinions due to personalized content brought by digital media (Borgesius et al., 2016). We argue that the primary concern of information cocoon research is the restrictions on individual's capability of striding across the cultural space.

However, prior research on information cocoon still has some shortcomings. First, the geometric nature of information cocoon has not been seriously considered in previous research. For example, there is a lack of geometric measurements for information cocoon. Second, prior studies on information cocoon are primarily limited to the study of political polarization (Bessi et al., 2016; Eady et al., 2019) and largely ignore information cocoons in cultural consumption and other non-political social scenarios. Third, there is a debate on the existence of information cocoons (Borgesius et al., 2016; Yuan & Ksiazek, 2015). Fourth, although social class plays a prominent role in cultural consumption (Bourdieu, 1984; Chan & Goldthorpe, 2005), its impact on the formation of information cocoon has not been well studied. To address these research gaps, the purpose of this study is twofold: First, to formally provide a theoretical framework of information cocoon by conceptualizing it as a cultural space; and second, to examine the existence of information cocoons as well its relation with social class.

The extensive use of mobile phones and computers provides us an ideal test ground to unobtrusively study the phenomenon of information cocoon. Due to their ubiquity and functionality, smartphones and computers are arising as one of the primary sensors of human behaviors (Lee & Cho, 2007; Yeykelis et al., 2014). On the one hand, people carry out a wide variety of tasks on digital devices, including reading books, listening to music, playing games, watching videos, searching for information, socializing, and so on. On the other hand, the possessions of digital devices reflect social status, personal characteristics, and other cultural meanings (Dittmar, 1991). For example, the wealth of people can be effectively inferred from their behaviors of mobile phone use (Blumenstock et al., 2015; Soto et al., 2011). In this study, we use three large-scale datasets to investigate the existence of information cocoon and its relationship with social class.

We claim that information cocoon could be better understood by conceptualizing the digital media as a cultural space and social class places restrictions on people's capability to move through the cultural space. By tracing the individual attention flow from one piece of information to another in the cultural space, we aim to provide a geometric framework of information cocoon and measure it with the word embedding models. In the following sections, we would first review the research on information cocoon, and argue that the concept of information cocoon provides us a geometric understanding of the cultural space. Second, we try to further clarify the theoretical underpinning between social class and information cocoon, and develop the research hypotheses. Third, we will introduce the methods of this study as well as the results. Last but not least, we would summarize the research findings and discuss the implications.

**The Geometric Nature of Information Cocoon**

Digital media provides researchers a new lens to investigate information cocoon since it documents our behavioral fingerprints. By observing peoples' media use behaviors, such as television viewing (Webster & Ksiazek, 2012), book purchasing (Shi et al., 2017), Facebook, Youtube, and Twitter use (Bessi et al., 2016; Eady et al., 2019), researchers can unobtrusively measure the human behaviors of cultural consumption. However, it is difficult to operationalize the abstract concepts, such as information cocoon, with behavioral data. Fortunately, the development of computational methods is changing this situation. Before introducing our research design and methodological standpoint, it is necessary to clarify the geometric nature of information cocoon.

Information cocoon is primarily a cultural space in which each individual's attention can flow from one kind of information to another. The terms of information cocoon, echo chamber, filter bubbles, and information enclaves imply that there is an enclosed space for each individual within the media environment. For example, in the book titled *Infotopia: How Many Minds Produce Knowledge,* Sunstein defines information cocoon as "communications universes in which we hear only what we choose and only what comforts and pleases us" (Sunstein, 2006, p.9). Sunstein further claims that his central interest has been in *aggregating knowledge across space* (Sunstein, 2006, p.122). However, the nature of information cocoon as a cultural space has not been seriously considered in later research. Part of the reason is that it is a high dimensional problem that is hard to think about or model. With the aid of behavioral data and computational methods, we can now trace the flow of collective attention to model the entire media environment as a cultural space, and locate the information cocoon for each individual within this cultural space.

The whole cultural space, as well as each individual's information cocoon, takes shape due to the interplay between media and users. Both institutions and users mutually construct the media environment. Webster (2011) proposes a structurational theory as a mechanism to explain the formation of audiences. On the one hand, as an essential kind of institution, media provide various genres of content to users. On the other hand, users allocate their attention to different categories of information. In addition to the media (institution) and users (agents), Webster argues that the public measures, such as advertising, audience rating, and recommendation systems, also play an essential role in the structurational process (Webster, 2011). The rise of digital media leads to a supply of massive information to users of diverse interests. The amount of information is beyond the audience's capability of information processing. As a result, most people choose to consume only the personalized information they are most interested in. And the public measures of personalized information further intensify this trend. That is to say, the personalized media environment compresses the range of individual movement in the cultural space and leads to the so-called phenomenon of information cocoons.

Given that information cocoons are a kind of compressed cultural space, it is necessary to study it from the geometric perspective. The ripe of the word embedding models provide a brand new method for us to conceptualize information cocoon as a geometric space. We propose that the cultural field could be better described with a vector space produced by word embedding models. To be specific, the word embedding models supply an efficient way to learn high-dimensional distributed vector representations that capture both syntactic and semantic relationships between words (Mikolov, Chen, et al., 2013; Mikolov, Sutskever, et al., 2013). Each word of the input text is positioned in the space in line with the context surrounding it. Thus, words of shared contexts are close to each other, while words of different contexts are far away from each other. It has been used to study various social and cultural issues, such as gender bias in language, the geometry of social class, and gender stereotypes in stories (Caliskan et al., 2017; Garg et al., 2018; Kozlowski et al., 2019; Xu et al., 2019).

Using the sequence of human behaviors as the input of word embedding models, we can construct the cultural space and observe the distribution of information within the space. The cultural space is a high dimensional geometric space, where each point within it could be precisely represented by a set of coordinates. Users selectively wander within the cultural space. By moving from one point to another, each individual weaves his or her own information cocoon. Thus, by tracing an individual's sequence of attention flow or mobility from one point to another, his or her information cocoon can be outlined. The size of an individual's information cocoon indicates his or her capability to stride over the cultural space. Further, we can also measure the distance from each person to a specific genre of content. Therefore, we can see which kind of information cocoon one person is stuck into. In this way, we provide a general geometric approach to investigate the existence of information cocoon.

Conceptually, the cultural space is an essential part of cultural field proposed by Bourdieu (1984). The cultural field is a social space in which individuals and objects are located in homologous positions relative to each other. Human beings live in a social space in which we can define and compare our behaviors from different social and cultural dimensions (Bourdieu, 1984, p.344). And they struggle for both material and symbolic resources in the social space. Bourdieu regularly portrays the cultural field geometrically using correspondence analysis. In his seminal book *Distinction*, Bourdieu visually presents the space of social positions (various of capitals) and the space of lifestyles (Bourdieu, 1984, p.128-129). Besides, he also proposes to study the relations between entities by measuring their distances (Bourdieu, 1984; Kozlowski et al., 2019). In this study, we describe the cultural field with the vector space produced by word embedding models. In all, conceptualizing information cocoon as a geometric space brings us back to the original idea of Bourdieu (1984) and other theorists (Sunstein, 2006; Jamieson & Cappella, 2008) of cultural consumption.

*The debate on information cocoon*

With the explosion of digital media, many scholars concern that information cocoon will be strengthened because people are provided with increasingly more personalized choices (Jamieson & Cappella, 2008; Pariser, 2011; Sunstein, 2006). To date, people can select the information or community that supports their existing opinions through selective exposure or algorithmic personalization. As a result, Anderson (2006) asserts that the so-called massive parallel culture will emerge and replace the hit-driven culture. In the mass media era, the popularity of media content plays a key role in peoples' media choice. The result is the winner takes all, and the corresponding culture of media consumption is regarded as the hit-driven culture. However, with the advent of digital media, Anderson (2006) proposed that the masses will be split into small interest tribes that do not have overlaps with each other. If this is the case, the massive parallel cultural will dominate the cultural consumption of digital media.

However, there are also different opinions. For example, Dubois and Blank (2018) contend that this kind of concern is overstated. They argue that the high-choice media environment also allows people to consume a wide variety of media. Their findings reveal that media diversity and political interest help reduce the probability of being caught in a partisan echo chamber. By measuring the echo chamber effect in multiple dimensions using questionnaires, they conclude that there is little evidence of partisan echo chamber. Similarly, Webster and his colleagues examined the TV market in the US, and their findings support the existence of the overlapping culture rather than the parallel culture (Webster & Ksiazek, 2012). Even though citizens have a preference for similar content on news outlets, they do not avoid disagreeable information (Valkenburg & Peter, 2013; Weeks et al., 2016).

There are also many methodological issues among the reasons accounting for the debate on the existence of information cocoon. For example, Dubois and Blank (2018) argue that people are living in multiple media environments. Therefore they suggest investigating how people interact with their entire media environment rather than a single platform of media. Also, prior research primarily relies on survey or experiment approaches to explore peoples' preference for like-minded content. As a result, the sample size is often limited, and only coarse-grained information could be collected. It is also problematic to rely on self-reported methods to measure information cocoon since it is difficult for people to recall whether and when they were exposed to different content or opinions (Prior, 2009). Furthermore, people may overestimate their capabilities of avoiding like-minded information, so the survey research of information cocoon is potentially flawed.

In all, the existence of information cocoon is still debatable. Different stages, different environments, and different media tend to affect the formation of information cocoon (Yuan & Ksiazek, 2015). Based on the literature above, we propose the main research question of this study:

RQ: Does information cocoon really exist in the daily use of digital media?

*Information cocoon and social class*

Social class is "the systematic and hierarchical distinction between persons and groups in social standing" (Kozlowski et al., 2019, p.3). As one of the most important concepts, there are multiple distinct dimensions of social class, such as wealth, occupational positions, education, esteem, and cultivated tastes. Marx (1977) argues that social class is determined by people's status or relationship in a specific production system and economic structure. In other words, the "social-structural position and relation to capital" is the basis of social class (Kozlowski et al., 2019, p.3). Weber (1958) asserts that esteem, such as social honor and prestige, plays a vital role in social class. He divides social class into three aspects, namely economic capital, political capital, and social capital (Weber, 1958). Besides, the cultivated taste is also an important element of social class. For example, Bourdieu interprets social class with the concept of habitus, which is a series of socially constructed dispositions or preferences that distinguished from other groups (Bourdieu, 1984).

Social class is one of the most important factors that limit people's capabilities to move in the cultural space. For example, in the seminal work of Bourdieu (1984), he claims that the pattern of cultural consumption is an expression of the structure of social class, which is regarded as the homology hypothesis. According to this hypothesis, the people of higher social class would prefer high or elite culture, and those of lower social class would consume popular or mass culture. Another explanation is the omnivore-univore hypothesis (Chan & Goldthorpe, 2005). The people who occupy advantaged positions of social class are cultural omnivores, while those of lower social class are cultural univores. On the one hand, the people of higher social class are more tolerant and open to different cultural styles; On the other hand, they still maintain certain cultural distinctions. Apparently, the relation between social structure and cultural consumption has become far more complex.

Social class plays a prominent role in our knowledge acquisition using mass media (Tichenor et al., 1970). The knowledge gap hypothesis suggests people of higher social class, especially for those highly educated, tend to access the information faster than those of lower socioeconomic status. Originated from the knowledge gap hypothesis, the term of digital divide describes a gap in terms of differential access and skills to technologies and information in the era of digital media (Compaine, 2001). Those of higher social and economic status know how to reap more benefits from the latest technology. Nowadays, as more people have access and skills to social media, researchers are more focused on the consumption of online content (Robinson et al., 2015). According to Prior's argument, people who consume more information may encounter more perspectives and arguments (Prior, 2007).

Taken together, the people of higher social class are more capable of moving within the cultural space than those of lower social class. Based on the above arguments, we propose the following hypothesis:

H1: People of higher social class have a larger possibility to escape from information cocoon than those of lower social class.

By measuring the cultural distances to different types of information, we may observe the footprints of digital distinction nowadays. In this paper, the cultural distance for each individual is defined as the geometric distance from one to each genre of content in the cultural space. As we have illustrated, we can locate the position of each user based on the content they consume. Similarly, based on all the content of a specific genre, we can also get the position for that genre of content. In this way, we can measure the geometric distance from each user to each genre of content.

Among all the genres of content, entertainment lies in the center of our cultural consumption in contemporary society. According to Neil Postman (1985), what we should fear is not that we become a captive culture just as George Orwell has warned us (Orwell, 1949), but we will become a trivial culture. As Aldous Huxley has argued in the book of Brave New World (Huxley, 1932), people are controlled by inflicting pleasure. Neil Postman (1985, p.13) asserts that our society has undergone a "shift from the magic of writing to the magic of electronics". In the era of digital media, taking TV as an example, the information is packaged as entertainment. Besides, cognitive resources for everyone is limited. The entertainment content might impede people's cognitive function if it occupies too many mental resources. Those who prefer reading news or science not only become more knowledgeable, but also get the ability to stride across the cultural space (Prior, 2005). In contrast, it is more difficult for those who consume the entertainment content a lot to escape from information cocoon. Based on the arguments above, we further propose the following hypothesis on the relation between information cocoon and the cultural distance to the content of entertainment:

H2: People who are farther away from the content of entertainment have a larger possibility to escape from information cocoon.

In the light of Bourdieu (1984), the habitus (i.e., internalized dispositions) or cultivated tastes mediate the impact of social class on cultural consumption. The distance to entertainment measures the cultural distance from each user to the entertainment content in the cultural space. An individual who is close to the entertainment content shall have a larger probability of preferring the entertainment content. That is to say, the distance to entertainment actually reveals people's cultivated taste in the entertainment content. If social class shapes our habitus or cultivated tastes, we also expect that social class can influence the distance to entertainment.

Further, according to omnivore-univore hypothesis, the people of higher social class are more open and appreciate different cultures (Chan & Goldthorpe, 2005; Peterson & Kern, 1996). In addition to being attracted by the popular content, such as the content of entertainment, they can still spend time on the other genres of content. As a result, the people of higher social class can better escape from the information cocoon driven by the tide of entertainment. Based on the arguments above, we propose the third hypothesis:

H3: People of higher social class have a larger distance to the content of entertainment than those of lower social class.

**Methods**

*Data*

We have collected three datasets of digital media use to study the geometric space of information cocoons. The first dataset contains 979 representative computer users in China, collected by China Internet Network Information Center. The user profile information, namely, gender, age, city, education level, job, and income, is collected through an online survey. There are 38,837,777 website browsing records within four weeks. There are fifteen categories of websites, including video, game, music, reading, news, office software, tool software, communication, forum, payment, shopping, travel, search, life service and real estate. These categories are labeled by five graduate students. We train these students until they achieve the accuracy of average 95.8% on a subset of the dataset.

The second dataset records 192,051,102 uses of smartphone applications of 100,000 users randomly selected in a major city of China within one month. In addition to online records, this dataset also contains offline locations of cell tower (longitude, latitude). The user profile data includes house price, gender, and age. There are eighteen predefined categories of application, including navigation, news, app store, real estate, search, payment, travel, game, weather service, shopping, communication, forum, webpage browsing, video, software management, email, reading, and music.

The third dataset documents 16,150,319 mobile reading behaviors from a random sample of 159,373 users in China within three months. The user profile is made up of (whether and how much a user paid for mobile reading). There are eighteen predefined categories of application, including navigation, news, app store, real estate, search, payment, travel, game, weather service, shopping, communication, forum, webpage browsing, video, software management, email, reading, and music.

These three datasets have their advantages separately and could be combined together to explore the relation between social class and information cocoon. The dataset of computer use has direct measurements of social class. The dataset of smartphone application use has large-scale user behaviors. The dataset of mobile reading is a specific app use and has records for three months.

*Constructing the geometric space*

We construct the geometric space using the word embedding models. To be specific, we employ the doc2vec model to embed users and content into a high dimensional space. The doc2vec method trained for variable-length text can perfectly match the variable-length time measure in mobile phone use (Le & Mikolov, 2014). Each user can be viewed as a document, and his or her sequence of behaviors over time can be viewed as words. Using the doc2vec model, we can build the geometric space of digital media, and get the coordinates of both users and the content browsed by them through mobile phone.

The input is the user id together with the sequence of content consumed by users. We train the doc2vec model using the Python package Gensim. The model parameters are chosen as follows: size=300 (vector length), min_count=1 (minimum frequency of words used in training), epoch=70 (number of training epochs over the training corpus). After the model is trained, both users and the content consumed by them are embedded into a high dimensional geometric space.

Figure 1 presents a 3-dimensional visualization of the geometric space for the dataset of mobile reading. We choose the top 20 similar books around 15 book genres. After the model is trained, we use the TSNE method for dimensionality reduction and visualization of the data. The content of the same genre (visualized with the same color and shape) tends to locate nearby.

### *Radius of information cocoon*

By embedding the information and users into a high dimensional space, we can investigate the geometric nature of information cocoons, especially the size or radius of information cocoon. We can measure the radius of information cocoon for each user with radius of gyration. The radius of gyration is widely used in the research of human mobility, and it typically measures a user's range of moving trajectory (Song et al., 2010). We can calculate the radius of gyration as following:

$$r_g = \sqrt{\frac{1}{L}\sum_{i=1}^{L}(r_i - r_{cm})^2}$$

where $r_i$ is the position at time i, L is the sum of recorded locations for an individual, $r_{cm}$ represents the center of mass of the trajectory: $r_{cm} = \frac{1}{L}\sum_{i=1}^{L} r_i$.

In Figure 1, we visualize the moving trajectories of two users. For user A, his mobile reading is limited to fiction and his moving radius is much smaller. For user B, he has a wider radius of mobile reading due to wider reading content.

Using the user vector trained with the doc2vec model as an approximation of the center of mass, we can calculate the radius of gyration ($M = 0.41$, $SD = 0.03$) in the dataset of computer use, the radius of gyration ($M = 0.48$, $SD = 0.03$) in the dataset of smartphone application use, and the radius of gyration ($M = 0.42$, $SD = 0.04$) in the dataset of mobile reading for each user of digital media.

*Range of information cocoon*

The range of information cocoon is the difference between the maximum distance and the minimum one except the entertainment content. The calculation of radius of gyration might drown an individual's desire to escape from information cocoon due to more browse of familiar content, whereas the range can capture the attempt. Also, different from the radius of information cocoon, we measure the range of information cocoon to distinguish entertainment and other categories. We measure the range of information cocoon in the dataset of computer use ($M$ = 0.18, $SD$ = 0.07), smartphone application use ($M$ = 0.14, $SD$ = 0.07), and mobile reading ($M$ = 0.14, $SD$ = 0.10).

*Cultural distance to genres*

We consider the distance from users to specific genres of content consumed by mobile phone users. The distance to entertainment is measured as: $-\frac{D_{max}^{entertainment} - D_{min}^{entertainment})}{D_{max}^{all} - D_{max}^{all}}$, where $D_{max}^{entertainment}$ and $D_{min}^{entertainment}$ represent the maximum and minimum distance to an individual in the genre of entertainment, and $D_{max}^{all}$ and $D_{min}^{all}$ represent the maximum and minimum distance to an individual in all the genres. For the data of computer and smartphone application use, we classify video, game and music as entertainment; while for the data of mobile reading, fiction, romance, and fantasy are considered as entertainment. In this way, we measure the distance to entertainment for the data of computer use ($M$ = -0.54, $SD$ = 0.27), smartphone application use ($M$ = -0.27, $SD$ = 0.20), and mobile reading ($M$ = -0.79, $SD$ = 0.30).

Based on the data, we divide the distance to entertainment into 5 ranks (from lowest 1 to highest 5).

### *The number of categories, relative entertainment preference*

The number of categories represents an individual's moving range, which is the baseline of the range of information cocoon. The number of category is calculated for the data of computer use ($M = 9.81$, $SD = 2.67$), smartphone application use ($M = 9.32$, $SD = 4.40$), and mobile reading ($M = 4.40$, $SD = 2.63$). Prior (2005) uses a measure called relative entertainment preference to indicate people's preference for entertainment consumption in media: $\frac{Time\ consumed\ in\ Entertainment}{Time\ consumed\ in\ All\ genres}$, which can be used as the baseline of the distance to entertainment. Relative entertainment preference is calculated for the data of computer use ($M = 0.05$, $SD = 0.07$), smartphone application use ($M = 0.06$, $SD = 0.11$), and mobile reading ($M = 0.79$, $SD = 0.33$).

### *Social class, gender, and age*

In the computer use dataset, job, education, income and city rank are taken as a proxy of social class. The job is categorized into 5 groups (unemployed, farmers, workers, clerks, managers). The education level is recorded into 6 groups (primary school and below, middle school, high school, Junior College, bachelor degree, master degree and above). The income is measured as an ordinal variable with 10 categories (0, 1~500, 501~1000, 1001~1500, 1501~2000, 2001~3000, 3001~5000, 5001~8000,

8001~12000, >12000). The city where they live is matched into 6 levels (from lowest 1 to highest 6).

In the smartphone application use dataset, we use the house price ($M = 33107.93$, $SD = 14972.62$) as a proxy to measure social class. As for house price, we estimate it by inferring an individual's home from the mobility records. Using the method for identifying origin-destination trips (Alexander et al., 2015; Dong et al., 2016; Xu et al., 2017), we extract the locations of massive anonymous users from the raw mobile phone data, and classify the locations into three types: home, work, and others. Finally, we match home location with the corresponding house price in that area.

While in mobile reading dataset, users' payment for reading (77.32% pay and 22.68% not) can be used as a proxy for social class.

The demographic variables of gender (female 22.4%, male 77.6%) and age ($M = 31.02$, $SD = 9.27$) in the dataset of computer use are included into the regression models as control variables. Similarly, the variables of gender (female 66.81%, male 19.07%, unknown 14.12%) and age ($M = 32.17$, $SD = 8.66$) in the dataset of smartphone application use are included into the regression models as control variables.

**Results**

*Most people live in information cocoons*

According to prior research, we formulate the research question about the existence of information cocoons. We can view a user's radius of gyration as his or her size of information cocoon. If the observed radius of gyration is significantly smaller than the expected value, we can conclude that the user is stuck into an information cocoon.

Given the observed value of radius of gyration, to test the statistical significance of the radius of gyration, we still need to know the expected values of the radius of gyration. To this end, we have to find a zero model that can generate expected values of the radius of gyration. We construct the zero models by randomly shuffling the sequence of content consumed by each user, and ensure that any two adjacent content is different and the length of a user's sequence is not changed. Then we train doc2vec models using the randomly shuffled sequences of content as the input, and calculate the expected values of radius of gyration. By repeating the random shuffling multiple times, we ensure that the generated distributions of the expected values are stable.

Figure 2 demonstrates the distribution of radius of gyration for both observed values and expected values for two datasets. The mean value of the observed distribution of radius of gyration is significantly smaller than the expected values for the dataset of computer use ($t(978) = -82.47, p < 0.001$), the dataset of smartphone app use ($t(99999) = -203.19, p < 0.001$) and the dataset of mobile reading ($t(159372) = -965.50, p < 0.001$). Therefore, the null hypothesis is rejected, and our findings support the existence of information cocoons in digital media, at least in the social settings of smartphone use.

*Social class and information cocoons*

We construct multiple linear regression models to test our hypotheses. The independent variable is the moving range, which measures the information cocoon effect. The reason why we do not use the variable of radius of gyration is that almost everyone is trapped in the information cocoon and there is no significant difference from this perspective. As shown in Table 1 in the dataset of computer use dataset, social class does not have a significant effect on radius of gyration. Therefore, we try to capture the possibility to escape from information cocoon with the variable moving range. The larger the range is, the smaller the information cocoon effect is.

The first hypothesis *H1* asserts that the people of higher social class have a larger moving range. In Table 2, the measurements of social class, such as job ($B = 0.03$, $p < 0.01$), education ($B = 0.02$, $p < 0.05$), and city rank ($B = 0.01$, $p < 0.05$), have a positive effect on the range for the dataset of smartphone application use; Similarly, house price ($B = 0.01$, $p < 0.01$), has a positive effect on the range for the dataset of smartphone application use; the payment on mobile reading ($B = 0.0002$, $p < 0.001$) also has a positive effect on the range for the dataset of mobile reading. Therefore, the hypothesis *H1* is supported by our datasets.

The second hypothesis *H2* is about the impact of distance to entertainment on the range of information cocoon. The distance to entertainment has a positive effect on the range of information cocoon for the dataset of smartphone application use ($B = 0.02$, $p < 0.001$), smartphone application use ($B = 0.02$, $p < 0.001$) and mobile reading ($B = 0.04$, $p < 0.001$). That is to say, if a user can stay farther away from the popular entertainment content, they can have a larger range of mobility in other genres in the cultural space, i.e., a smaller information cocoon effect. Therefore, the hypothesis *H2* is well supported.

The third hypothesis *H3* further asserts that those of higher social class have a larger distance to entertainment than those of lower social class. In Table 3, the regression model has revealed that the measurements of social class, such as education ($B = 0.15$, $p < 0.01$), have a positive effect on the distance to entertainment for the dataset of computer use. Likewise, the measurement of social class, house price ($B = 0.08$, $p < 0.001$) has a positive effect on the distance to entertainment for the dataset of smartphone application use. Similarly, the payment on mobile reading ($B = 0.10$, $p < 0.001$) also has a positive effect on the distance to entertainment for the dataset of mobile reading. Therefore, the hypothesis *H3* is supported by our datasets.

In addition, for the dataset of smartphone application use, we have also controlled the influence of gender and age. Compared with females, males tend to have a larger distance to the content of entertainment ($B = 0.06$, $p < 0.001$). The older people have a larger distance to the content of entertainment than the young people ($B = -0.01$, $p < 0.001$).

**Discussion and Conclusion**

Our methodological contribution is to conceptualize information cocoon as a cultural space, which highlights the geometric nature of information cocoon. Thanks to the digital traces documented by digital media, we can measure the sequence of human behaviors at a large scale and in fine-grained resolution. Using word embedding models, we can embed both users and the content consumed by them into a high-dimensional geometric space (Kozlowski et al., 2019). Thus, we can better represent the positions of users and content with distributed vectors, quantify the similarity of social meanings for the content, and observe how users traverse this cultural space.

The key findings of this study suggest that the phenomenon of information cocoons widely exists in the daily use of digital media. As a result, each individual will become increasingly more ignorant and pay less attention to public issues. Also, along with the deterioration of the cultivated tastes, the mass culture will drive out the high culture out of circulation. In the words of Postman (1985), the culture has become a burlesque. The ubiquitous existence of information cocoons implies that Aldous Huxley's worry about the trivial culture has come true (Huxley, 1932). Therefore, it becomes even more difficult to aggregate local information into global knowledge across the cultural space. The ideal of many minds working together to produce knowledge actually becomes a utopia (Sunstein, 2006).

The finding about the impact of social class supports the omnivore-univore hypothesis (Chan & Goldthorpe, 2005): the people of higher social class tend to be omnivores who are more open to different cultural content, while the people of lower social class tend to be univores who have a much smaller range of cultural consumption. Besides, our findings reveal that the people of lower social class have a shorter distance to the entertainment content, which further intensifies the problem of information cocoon. Taken together, the vulnerable groups are more likely to be caught in information cocoons. The disadvantages in accessing and using digital media prevent the vulnerable groups from acquiring knowledge better, which may further widen social inequality (Compaine, 2001; Prior, 2007; Robinson et al., 2015).

Further, our findings indicate that the information cocoon in computer use tends to be weaker than that in smartphone use. We argue that mobile phone constrains the flexibility of our media use compared with personal computers. First, the screen size of mobile phones is much smaller than personal computers; Second, the degree of customization is much stronger in the smartphone applications than that of personal computers; Third, its more flexible to get information using personal computers than smartphones. For example, it is more convenient to search online using computers than using smartphones. Additionally, our finding suggests that the impact of social class tends to be weaker in computer use relative to smartphone use. One possible explanation lies in the fact that the social class of computer users tends to be more homogeneous than that of smartphone users.

It is also necessary to note the limitations of this research. Although we have employed three different types of datasets to measure the information cocoon, our empirical observations are limited to the digital media users of China. Future research should examine the existence of information cocoon as well as its relationship with social class in other cultural contexts.

In all, social class impedes the flow of individual attention. Our findings support the existence of information cocoon in our daily use of digital media. Conceptualizing the digital media as a cultural space, we theorize the daily use of digital media as a social process during which users move within the cultural space. Unfortunately, our capability of traversing the cultural space over time tends to be limited, especially for those of lower social class who tend to be trapped in the content of entertainment.

**References**

Alexander, L., Jiang, S., Murga, M., & González, M. C. (2015). Origin–destination trips by purpose and time of day inferred from mobile phone data. Transportation Research Part c: Emerging Technologies, 58, 240–250. doi: 10.1016/j.trc.2015.02.018

Anderson, C. (2006). The long tail: Why the future of business is selling less of more. New York: Hachette Books.

Bessi, A., Zollo, F., Del Vicario, M., Puliga, M., Scala, A., Caldarelli, G., Uzzi, B., & Quattrociocchi, W. (2016). Users polarization on Facebook and Youtube. PLOS ONE, 11(8), e0159641. doi: 10.1371/journal.pone.0159641

Blumenstock, J., Cadamuro, G., & On, R. (2015). Predicting poverty and wealth from mobile phone metadata. Science, 350(6264), 1073–1076. doi: 10.1126/science.aac4420

Borgesius, F. J. Z., Trilling, D., Möller, J., Bodó, B., De Vreese, C. H., & Helberger, N. (2016). Should we worry about filter bubbles? Internet Policy Review. Journal on Internet Regulation, 5(1), 1-16. doi: 10.14763/2016.1.401

Bourdieu, P. (1984). Distinction: A critique of the social judgement of taste. Cambridge, Mass.: Harvard University Press.

Caliskan, A., Bryson, J. J., & Narayanan, A. (2017). Semantics derived automatically from language corpora contain human-like biases. Science, 356(6334), 183–186. doi: 10.1126/science.aal4230

Chan, T. W., & Goldthorpe, J. H. (2005). The social stratification of theatre, dance and cinema attendance. Cultural Trends, 14(3), 193–212. doi: 10.1080/09548960500436774

Compaine, B. M. (2001). The digital divide: Facing a crisis or creating a myth? Cambridge, Mass.: MIT Press.

Dittmar, H. (1991). Meanings of material possessions as reflections of identity: Gender and social-marterial position in society. Journal of Social Behavior and Personality, 6(6), 165-186.

Dong, L., Li, R., Zhang, J., & Di, Z. (2016). Population-weighted efficiency in transportation networks. Scientific reports, 6, 26377. doi: 10.1038/srep26377

Dubois, E., & Blank, G. (2018). The echo chamber is overstated: The moderating effect of political interest and diverse media. Information, Communication & Society, 21(5), 729–745. doi: 10.1080/1369118X.2018.1428656

Eady, G., Nagler, J., Guess, A., Zilinsky, J., & Tucker, J. A. (2019). How many people live in political bubbles on social media? Evidence from linked survey and Twitter data. SAGE Open, 9(1), 215824401983270. doi: 10.1177/2158244019832705

Garg, N., Schiebinger, L., Jurafsky, D., & Zou, J. (2018). Word embeddings quantify 100 years of gender and ethnic stereotypes. Proceedings of the National Academy of Sciences, 115(16), E3635–E3644. doi: 10.1073/pnas.1720347115

Huxley, A. (1932). Brave new world. London: Chatto & Windus.

Jamieson, K. H., & Cappella, J. N. (2008). Echo chamber: Rush Limbaugh and the conservative media establishment. Oxford: Oxford University Press.

Kozlowski, A. C., Taddy, M., & Evans, J. A. (2019). The Geometry of culture: Analyzing the meanings of class through word embeddings. American Sociological Review, 84(5), 905–949. doi: 10.1177/0003122419877135

Le, Q., & Mikolov, T. (2014). Distributed representations of sentences and documents. International Conference on Machine Learning, 1188–1196. 22-24 June 2014, Bejing, China.

Lee, Y., & Cho, S.-B. (2007). Extracting meaningful contexts from mobile life log. International Conference on Intelligent Data Engineering and Automated Learning, 750–759. December 16-19 2017, Birmingham, UK. doi: 10.1007/978-3-540-77226-2_75

Marx, K. (1977). Capital, Vol. 1. New York: Vintage Books.

Mikolov, T., Chen, K., Corrado, G., & Dean, J. (2013). Efficient estimation of word representations in vector space. arXiv:1301.3781.

Mikolov, T., Sutskever, I., Chen, K., Corrado, G. S., & Dean, J. (2013). Distributed representations of words and phrases and their compositionality. Proceedings of the 26th International Conference on Neural Information Processing Systems -

Volume 2. 5-10 December 2013, Red Hook, NY, USA.

Orwell, G. (1949). Nineteen Eighty-Four: (1984). New York: Harcourt.

Pariser, E. (2011). The filter bubble: What the Internet is hiding from you. London: Penguin Press.

Peterson, R. A., & Kern, R. M. (1996). Changing highbrow taste: From snob to omnivore. American Sociological Review, 61(5), 900–907. doi: 10.1111/j.1540-5907.2005.00143.x

Postman, N. (1985). Amusing ourselves to death: Public discourse in the age of show business. New York: Viking Press.

Prior, M. (2005). News vs. Entertainment: How increasing media choice widens gaps in political knowledge and turnout. American Journal of Political Science, 49(3), 577–592. doi: 10.1111/j.1540-5907.2005.00143.x

Prior, M. (2007). Post-broadcast democracy: How media choice increases inequality in political involvement and polarizes elections. Cambridge: Cambridge University Press.

Prior, M. (2009). The immensely inflated news audience: Assessing bias in self-reported news exposure. Public Opinion Quarterly, 73(1), 130–143. doi: 10.1093/poq/nfp002

Qian, X., Feng, H., Zhao, G., & Mei, T. (2013). Personalized recommendation combining user interest and social circle. IEEE Transactions on Knowledge and Data Engineering, 26(7), 1763–1777. doi: 10.1109/TKDE.2013.168

Robinson, L., Cotten, S. R., Ono, H., Quan-Haase, A., Mesch, G., Chen, W., Schulz, J., Hale, T. M., & Stern, M. J. (2015). Digital inequalities and why they matter. Information, Communication & Society, 18(5), 569–582. doi: 10.1080/1369118X.2015.1012532

Shi, F., Shi, Y., Dokshin, F. A., Evans, J. A., & Macy, M. W. (2017). Millions of online book co-purchases reveal partisan differences in the consumption of science. Nature Human Behaviour, 1(4), 0079.

Song, C., Koren, T., Wang, P., & Barabási, A.-L. (2010). Modelling the scaling properties of human mobility. Nature Physics, 6(10), 818-823. doi:

10.1038/nphys1760

Soto, V., Frias-Martinez, V., Virseda, J., & Frias-Martinez, E. (2011). Prediction of socioeconomic levels using cell phone records. In: Konstan, J., Conejo Muñoz, R., Marzo, J.L., Oliver, N. (Eds.). International Conference on User Modeling, Adaptation, and Personalization, 377–388. 11-15 July, 2011. Girona, Spain. Berlin: Springer. doi: 10.1007/978-3-642-22362-4_35

Sunstein, C. R. (2006). Infotopia: How many minds produce knowledge. Oxford: Oxford University Press.

Tichenor, P. J., Donohue, G. A., & Olien, C. N. (1970). Mass media flow and differential growth in knowledge. Public Opinion Quarterly, 34(2), 159–170. doi: 10.1086/267786

Valkenburg, P. M., & Peter, J. (2013). The differential susceptibility to media effects model. Journal of Communication, 63(2), 221–243. doi: 10.1111/jcom.12024

Weber, M. (1958). The protestant ethic and the spirit of capitalism. New York: Charles Scribner's Sons.

Webster, J. G. (2011). The duality of media: A structurational theory of public attention. Communication Theory, 21(1), 43–66.

Webster, J. G., & Ksiazek, T. B. (2012). The dynamics of audience fragmentation: Public attention in an age of digital media. Journal of Communication, 62(1), 39–56. doi: 10.1111/j.1460-2466.2011.01616.x

Weeks, B. E., Ksiazek, T. B., & Holbert, R. L. (2016). Partisan enclaves or shared media experiences? A network approach to understanding citizens' political news environments. Journal of Broadcasting & Electronic Media, 60(2), 248–268. doi: 10.1080/08838151.2016.1164170

Xu, H., Zhang, Z., Wu, L., & Wang, C.-J. (2019). The Cinderella Complex: Word embeddings reveal gender stereotypes in movies and books. PlOS ONE, 14(11): e0225385. doi: 10.1371/journal.pone.0225385

Xu, Y., Li, R., Jiang, S., Zhang, J., & González, M. C. (2017). Clearer skies in Beijing– revealing the impacts of traffic on the modeling of air quality. The 96th Annual Meeting of Transportation Research Board. 8-12 January 2017. Washington DC,

USA.

Yeykelis, L., Cummings, J. J., & Reeves, B. (2014). Multitasking on a single device: Arousal and the frequency, anticipation, and prediction of switching between media content on a computer. Journal of Communication, 64(1), 167–192. doi: 10.1111/jcom.12070

Yuan, E., & Ksiazek, T. (2015). A network analytic approach to audience behavior and market structure: The case of China and the United States. Mass Communication and Society, 18(1), 58–78. doi: 10.1080/15205436.2013.879667

Figure 1. The Visualization of Cultural Space.

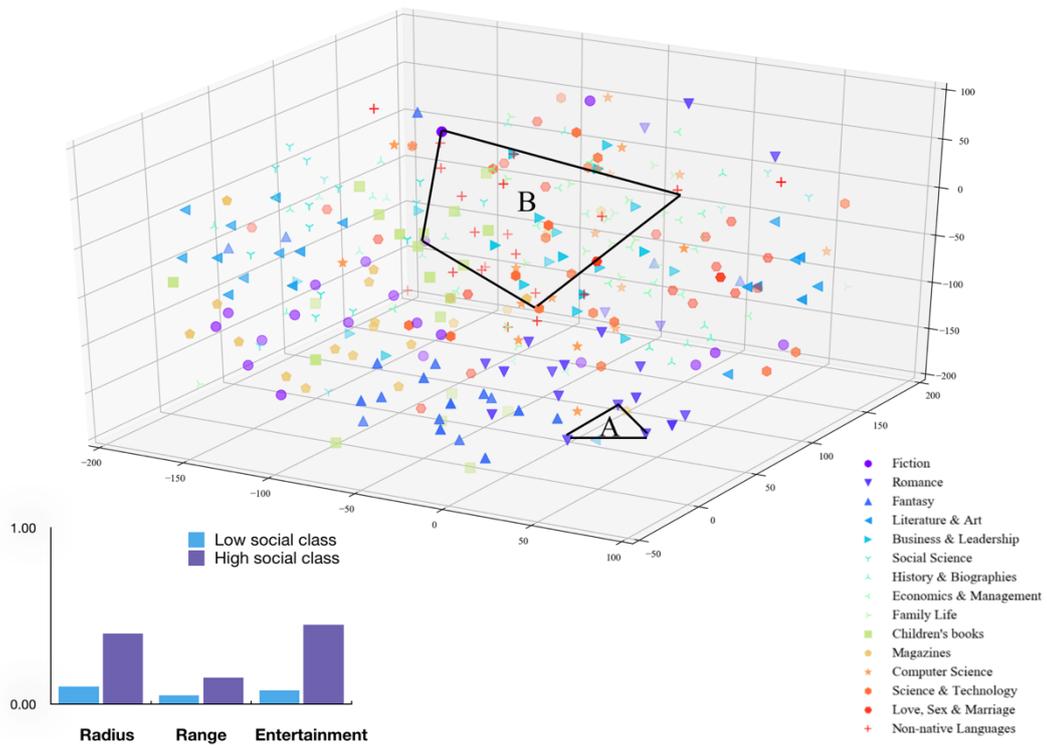

Figure 2. Most people Live in Information Cocoons.

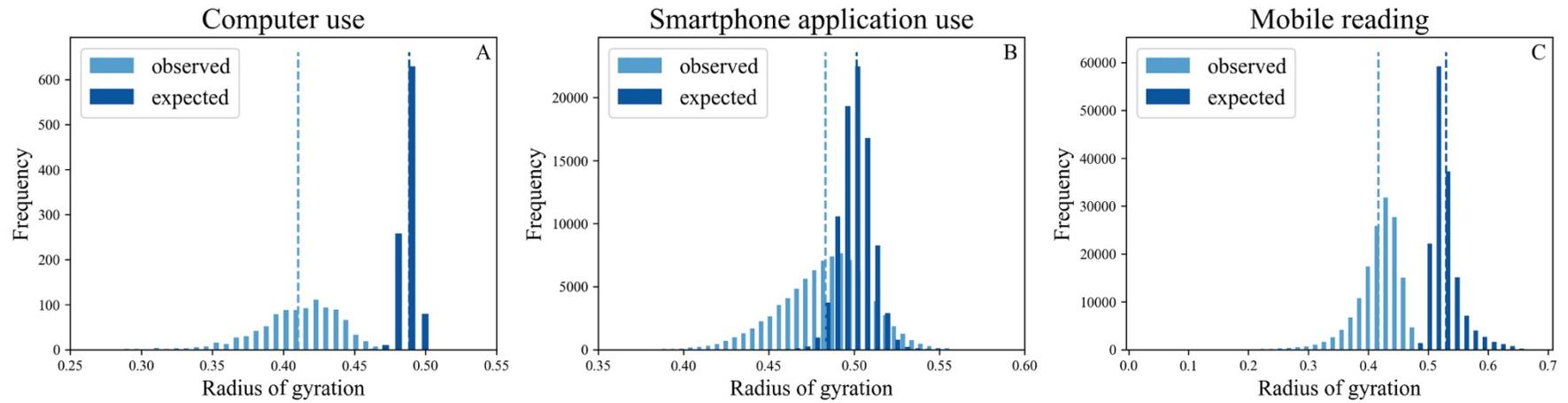

Table 1. OLS Regression Models of Radius of Information Cocoon.

|  | Model1: Mobile Reading | | Model2: Smartphone Use | | Model3: Computer Use | |
|---|---|---|---|---|---|---|
|  | Measure | B (SE) | Measure | B (SE) | Measure | B (SE) |
| Social Class | Payment | 0.01*** | House Price (Log) | 0.004*** | Job | -0.02 |
|  |  | (0.001) |  | (0.001) |  | (0.014) |
|  |  |  |  |  | Education | -0.03 |
|  |  |  |  |  |  | (0.033) |
|  |  |  |  |  | Income | -0.01 |
|  |  |  |  |  |  | (0.007) |
|  |  |  |  |  | City Rank | -0.01 |
|  |  |  |  |  |  | (0.007) |
| Control Variables | Category Num | 0.01*** | Category Num | 0.02*** | Category Num | 0.04*** |
|  |  | (0.000) |  | (0.000) |  | (0.004) |
|  |  |  | Gender | -0.01*** | Gender | 0.00 |
|  |  |  |  | (0.001) |  | (0.026) |
|  |  |  | Age | 0.0002** | Age | -0.00 |
|  |  |  |  | (0.000) |  | (0.001) |
| N | 159373 | | 85879 | | 731 | |
| $R^2$ | 0.06 | | 0.56 | | 0.14 | |

*Note.* * P < 0.05; ** P < 0.01; *** P < 0.001.

Table 2. OLS Regression Models of Range of Information Cocoon.

| | Model1: Mobile Reading | | Model2: Smartphone Use | | Model3: Computer Use | |
|---|---|---|---|---|---|---|
| | Measure | B (SE) | Measure | B (SE) | Measure | B (SE) |
| Social Class | Payment | 0.0002*** | House Price (Log) | 0.01** | Job | 0.03** |
| | | (0.000) | | (0.003) | | (0.012) |
| | | | | | Education | 0.02* |
| | | | | | | (0.010) |
| | | | | | Income | 0.00 |
| | | | | | | (0.006) |
| | | | | | City Rank | 0.01* |
| | | | | | | (0.006) |
| Control Variables | Category Num | 0.07*** | Category Num | 0.04*** | Category Num | 0.06*** |
| | | (0.000) | | (0.000) | | (0.003) |
| | | | Gender | -0.01** | Gender | 0.01 |
| | | | | (0.002) | | (0.022) |
| | | | Age | 0.001** | Age | 0.002 |
| | | | | (0.000) | | (0.001) |
| N | 159373 | | 85879 | | 731 | |
| $R^2$ | 0.46 | | 0.32 | | 0.38 | |

Note. * P < 0.05; ** P < 0.01; *** P < 0.001.

Table 3. OLS Regression Models of Distance to Entertainment.

|  | Model1: Mobile Reading | | Model2: Smartphone Use | | Model3: Computer Use | |
|---|---|---|---|---|---|---|
|  | Measure | B (SE) | Measure | B (SE) | Measure | B (SE) |
| Social Class | Payment | 0.10*** | House Price (Log) | 0.08*** | Job | 0.05 |
|  |  | (0.007) |  | (0.015) |  | (0.03) |
|  |  |  |  |  | Education | 0.15** |
|  |  |  |  |  |  | (0.053) |
|  |  |  |  |  | Income | 0.01 |
|  |  |  |  |  |  | (0.033) |
|  |  |  |  |  | City Rank | 0.02 |
|  |  |  |  |  |  | (0.03) |
| Control Variables | Relative Entertainment | -0.91*** | Relative Entertainment | -0.65*** | Relative Entertainment | -1.80** |
|  |  | (0.009) |  | (0.043) |  | (0.70) |
|  |  |  | Gender | 0.06*** | Gender | 0.04 |
|  |  |  |  | (0.012) |  | (0.119) |
|  |  |  | Age | -0.01*** | Age | 0.005 |
|  |  |  |  | (0.001) |  | (0.006) |
| N | 159160 | | 82835 | | 731 | |
| $R^2$ | 0.06 | | 0.004 | | 0.04 | |

*Note.* * P < 0.05; ** P < 0.01; *** P < 0.001.